\documentclass[sigconf, nonacm]{acmart}
\renewcommand\footnotetextcopyrightpermission[1]{}

\usepackage{pgf}
\usepackage{tikz}
\usetikzlibrary{arrows,automata}
\usetikzlibrary{shapes.geometric}
\usetikzlibrary{positioning}

\usepackage{balance}
\usepackage{algorithm}
\usepackage{algpseudocode}
\usepackage{xspace}
\usepackage{comment}
\usepackage{graphicx}
\usepackage{booktabs}
\usepackage[binary-units=true]{siunitx}
\usepackage{url}
\usepackage{amsmath}
\usepackage{multirow}

\usepackage{enumitem}
\setitemize{noitemsep,topsep=0pt,parsep=0pt,partopsep=0pt,itemindent=0em,leftmargin=1em}
\setenumerate{noitemsep,topsep=0pt,parsep=0pt,partopsep=0pt}
\setdescription{itemsep=0pt,topsep=0pt,parsep=0pt,partopsep=0pt,itemindent=0em,leftmargin=0.5em}

\microtypecontext{spacing=nonfrench}

\newcommand{\torchserve}{TorchServe\xspace}

\newcommand{\ms}[1]{\SI{#1}{\milli\second}}

\iftrue
\newcommand{\ankit}[1]{{\color{red}Ankit: #1} \newline}
\newcommand{\amar}[1]{\textcolor{blue}{Amar:#1}}
\newcommand{\ryan}[1]{\textcolor{purple}{Ryan:#1}}
\newcommand{\deepak}[1]{\textcolor{orange}{Deepak: #1}}
\else
\newcommand{\ankit}[1]{}
\newcommand{\amar}[1]{}
\newcommand{\ryan}[1]{}
\newcommand{\deepak}[1]{}
\fi

\newcommand{\itb}[1]{$\langle{}$#1$\rangle{}$}
\newcommand{\itbitb}{$\langle{}i$, $t$, $b\rangle{}$}
\newcommand{\itbTB}{$\langle{}T$, $B\rangle{}$}
\newcommand{\itbtb}{$\langle{}t$, $b\rangle{}$}

\newcommand{\sys}{Packrat\xspace}
\newcommand{\sysname}{Packrat\xspace}

\newcommand\vldbpagestyle{plain}

\begin{document}

\date{}

\title{\sys{}: Automatic Reconfiguration for Latency Minimization in CPU-based DNN Serving}

\author{Ankit Bhardwaj}
\authornote{Work done as a Microsoft Research Intern in Project Fiddle.}
\affiliation{%
  \institution{University of Utah}
  \city{}
  \country{}
}
\email{ankitb@cs.utah.edu}

\author{Amar Phanishayee}
\affiliation{%
  \institution{Microsoft Research}
  \city{}
  \country{}
}
\email{amar@microsoft.com}

\author{Deepak Narayanan}
\authornote{Work done when the author was at Microsoft.}
\affiliation{
  \institution{NVIDIA}
  \city{}
  \country{}
}
\email{dnarayanan@nvidia.com}

\author{Mihail Tarta}
\affiliation{%
  \institution{Microsoft}
  \city{}
  \country{}
}
\email{mtarta@microsoft.com}

\author{Ryan Stutsman}
\affiliation{%
  \institution{University of Utah}
  \city{}
  \country{}
}
\email{stutsman@cs.utah.edu}

\begin{abstract}
In this paper, we investigate how to push the performance limits of serving Deep Neural Network (DNN) models on CPU-based servers.  
Specifically, we observe that while intra-operator parallelism across multiple threads is an effective way to reduce inference latency, it provides diminishing returns.  
Our primary insight is that instead of running a single instance of a model with all available threads on a server, running multiple instances each with smaller batch sizes and fewer threads for intra-op parallelism can provide lower inference latency.  
However, the right configuration is hard to determine manually since it is workload- (DNN model and batch size used by the serving system) and deployment-dependent (number of CPU cores on server).  
We present \sys, a new serving system for online inference that given a model and batch size ($B$) algorithmically picks the optimal number of instances ($i$), the number of threads each should be allocated ($t$), and the batch sizes each should operate on ($b$) that minimizes latency.
\sys is built as an extension to TorchServe and supports online reconfigurations to avoid serving downtime.
Averaged across a range of batch sizes, \sys{} improves inference latency by 1.43$\times$~to~1.83$\times$ on a range of commonly used DNNs.

\end{abstract}

\maketitle

\pagestyle{\vldbpagestyle}

\section{Introduction}
\label{sec:introduction}

Deep Neural Network (DNN) serving is an increasingly important datacenter workload.
DNN serving systems are often used in online services like image and video analytics, speech transcription, text and code completion, chatbots, and more.
In these settings, requests arrive continuously and must be served in real time; thus, serving systems must handle high request rates efficiently and with low response latency.

There are many DNN serving systems available today, including TensorFlow Serving~\cite{tensorflow_serving}, \torchserve~\cite{torchserve}, and Triton~\cite{triton}.
These systems are designed to use both CPUs and GPUs to execute DNN model inference.
GPUs generally provide better throughput than CPUs, but they are often more expensive and power-hungry. They also end up underutilized for inference workloads~\cite{folded-cnns}.
Recent CPU advances, like high core counts (56~to~64 cores are common today~\cite{56-cores,64-cores})
and specialized instructions that support lower numerical precision multiplications with higher precision accumulates (AVX-512~\cite{avx512}, AMX~\cite{amx}), improve inference performance.
Every cloud server comes equipped with such multi-core CPUs and many product groups at large companies already own large fleets of such servers that are now used for CPU-based serving (including at this large cloud provider).

\begin{figure}
    \centering \includegraphics[width=1.0\linewidth]{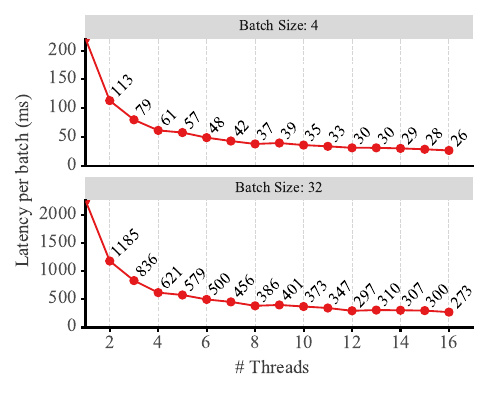}
    \caption{ResNet-50 inference latencies with different number of threads~($T$) for intra-operator parallelism with batch-size~$B$~=~4~and~32. Increasing the number of threads beyond a certain point provides diminishing returns,
        and this point of diminishing returns is model and batch size dependent.}
    \label{fig:resnet-latency-8}
\end{figure}

In this paper, we push the performance limits of serving DNN models on a single multicore CPU-based server.
Serving systems like Triton and \torchserve provide useful features like request handling, adaptive batching of inference requests, and multi-model serving.
One important technique to improve the latency of DNN model serving is intra-operator parallelism~\cite{shoeybi2019megatron}, where a single operator is split and run using multiple threads.
Figure~\ref{fig:resnet-latency-8} shows inference runtimes of a ResNet-50 model versus number of threads used for intra-operator parallelism (up to 16) on a single socket of a CPU-based server.
We show results for two batch sizes (4 and 32), highlighting the improvement on latency due to intra-operator parallelism.
We observe that increasing the number of threads results in diminishing returns; this observation is consistent across batch sizes and other models as well (\S\ref{sec:iop}).

In this paper, we present \sysname{}, an optimized CPU-based serving system for online inference that automatically determines the number of threads that need to be allocated to model instances to minimize inference latency.
\sysname{} is motivated by the following key insight: \textit{instead of running a single instance of a model with all available threads (the default for systems like TorchServe), running multiple instances each with smaller batch sizes and fewer threads for intra-op parallelism can provide lower inference latency}.
For example, Figure~\ref{fig:resnet-latency-8} shows that using a total of 16 threads ($T$) for a ResNet-50 model with a batch of size 32 ($B$) is sub-optimal (latency of 273~ms).
To sidestep the diminishing benefits from intra-op parallelism, a user might try to create one model instance per core and configure their workload to split each batch across the available threads, but this does not minimize latency either.
Instead, these measurements suggest that
running 8 inference instances ($i=8$) with 2 threads ($t=2$) each serving a batch of 4 items ($b=4$) could lower the latency of the entire batch over either of these configurations (to 113~ms, a $2.4\times$ speedup). In short, neither maximizing intra-op parallelism nor maximizing parallelism across model instances results in the best inference latency.

In the general case, determining the optimal configuration of $\langle$instances, threads, batch$\rangle$ (or \itbitb{} for short) is challenging because it is workload- and deployment-specific. The optimal configuration depends on the specific model being served, input dimensions like the batch size (which is itself dependent on the request arrival rate), and the hardware (e.g., number of cores, memory bandwidth, etc.).
Furthermore, even if there were a hypothetical oracle that could provide the optimal \itbitb{} configuration, a user would still have to manually recognize when to change configurations and then reconfigure existing serving systems while specifying thread-core affinities appropriately.

\sysname{} uses a novel algorithm to dynamically determine the optimal \itbitb{} configuration for models on individual servers given a batch of inputs for the model.
It does this automatically using a small amount of targeted profiling; from this limited profiling information, it formulates \itbitb{} configurations that are expected to optimize average batch latency for different batch sizes by solving a 2-dimensional knapsack problem using dynamic programming.
This lets \sys{} quickly find configurations that balance intra-op latency with multi-instance execution without the need for user input and without impractically profiling \emph{all possible} configuration combinations.
Combined with its mechanism of transitioning between configurations,
this lets \sys{} dynamically reconfigure model instances and threads used for inference, entirely online, so as to optimize inference latency as workloads change.
We evaluate \sysname{} on a single server running \torchserve.
Over several models, we show that \sys{} improves inference latency and throughput over the baseline approach that maximizes intra-op parallelism
by 1.43$\times$~to~1.83$\times$ averaged over a range of batch sizes.
\sys{} code will be made public with the final version of the paper.


\section{Background and Motivation}
\label{sec:background}
DNN inference involves executing a single forward pass of the model for each inference input.
The forward pass consists of a sequence of operations like matrix multiplications,
convolutions, vector operations, and activation functions that are executed in a specific order.
Each request incurs some overhead including data transformations, memory allocations, and data copying.
These overheads can be amortized by batching multiple inference inputs and
executing them in one forward pass, improving arithmetic intensity and overall performance of the system.

For a request batch, each operator processes a whole input batch to produce
a batched output, instead of going through individual operators input-by-input.
Typical production DNN serving systems like Triton~\cite{triton} and \torchserve~\cite{torchserve} support batched execution.
Many of these systems also support adaptive batching for online inference; for example, if a user-configured number of input items have not arrived within a timeout interval, they can send all available items as a batch of inputs for inference right away and not incur further queuing latency.

\subsection{Hardware-level Parallelism}
\label{sec:hardware-parallelism}

Modern server-class CPUs have 10s to 100s of cores~\cite{server-processors}.
For highly parallelizable operations like matrix multiplications,
many cores can execute the operation in parallel to reduce overall execution time.
In addition to multicore parallelism, each core can also execute
instructions in parallel, called instruction-level parallelism (ILP).
ILP is achieved using out-of-order execution~\cite{arch-book} of
instructions.

Modern CPUs also support SIMD instructions, which execute the same
instruction on multiple data elements in parallel.
The \texttt{AVX2} and \texttt{AVX512} instruction sets execute one instruction
on 256 and 512 bits in parallel, respectively.
For example, \texttt{AVX2} can execute 8~single-precision~(32-bit) floating-point operations in
parallel, and \texttt{AVX512} can execute 16~single-precision floating-point operations
in parallel.
Additionally
, modern CPUs have multiple execution units, such as integer,
floating-point, and vector units, which execute multiple
instructions in parallel.
Fused multiply-add (FMA) instructions~\cite{fma} can execute multiply and
add operations with one instruction, which can further improve
performance.

Optimized libraries like OpenBLAS~\cite{openblas} and Intel MKL~\cite{mkl} use advanced instruction sets like fused \texttt{AVX2} and \texttt{AVX512} for good performance.
Most optimized machine learning frameworks like PyTorch~\cite{pytorch} and
TensorFlow~\cite{tensorflow} use these optimized libraries internally.

Next, we focus on one optimization that is critical for improving inference latency: intra-operator parallelism.

\subsection{Intra-Op Parallelism}
\label{sec:iop}
As described earlier, each inference
involves executing a sequence of operations like matrix multiplication,
convolutions, or activation functions with vector operations in a specific order.
Each operation (e.g.,\ a single matrix multiply)
can be broken up and executed in parallel across multiple cores.
This is called \textsl{intra-op parallelism} because operators for a single input's inference (or a batch of them) are executed in a parallel fashion.
Depending on the implementation, intra-op parallelism is realized through
\texttt{OpenMP}~\cite{openmp} or using MKL threads. %
Optimized libraries like Intel MKL internally use vector
instructions to improve performance~(\S\ref{sec:hardware-parallelism}).
By default, \texttt{OpenMP} matches the number of threads it uses to the number of physical cores available on the machine when executing parallel code. However, DNN frameworks like
PyTorch and TensorFlow also allow the user to specify the number of threads to use.

\begin{figure*}
    \centering
    \includegraphics[width = 1.0\textwidth]{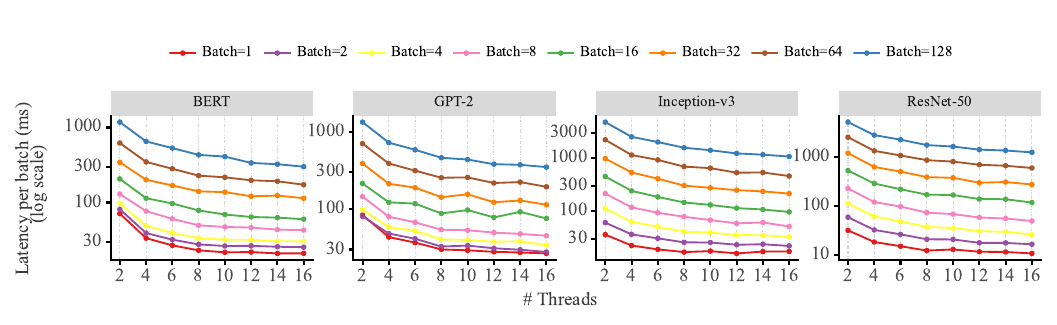}
    \caption{Intra-operator parallelism offers diminishing returns after scaling beyond a certain the number of threads.  Though the exact point and magnitude of diminishing returns may differ, this trend is consistent across different models and batch sizes.}
    \label{fig:thread-sweep}
\end{figure*}

To understand the impact of intra-op parallelism on inference performance, we execute inference for different models while sweeping through different batch sizes and number of threads.
Figure~\ref{fig:thread-sweep} shows this for four different models: ResNet-50~\cite{resnet}, Inception-v3~\cite{inception}, GPT-2~\cite{gpt2}, and BERT~\cite{bert}.
We find that for all these models, intra-op parallelism improves inference throughput and latency, but scaling the number of threads assigned to intra-op parallelism provide diminishing returns.
While this general trend is visible in Figure~\ref{fig:thread-sweep} across models and batch sizes, we focus on a single model
to illustrate the nuanced differences.

Figure~\ref{fig:resnet-latency-8} shows the results for ResNet-50 on a single serving instance while varying threads for intra-op parallelism when executing on a batch size of 4 and 32.
As an example, for a batch size of 4, increasing the number of threads for intra-op parallelism from 2 to 4 improves latency by $1.85\times$, but from 8 to 16 results in a $1.4\times$ improvement.
Similarly, for batch size 32, going from 2 to 4 threads improves latency by $1.9\times$, but going from 8 to 16 improves latency only by $1.4\times$.

Fortunately, serving systems like TorchServe allow users the flexibility to specify the number of threads assigned to serving a model instance.
For example, a user might allocate all threads on a server to process individual input batches using intra-op parallelism, with the hope of improving per-batch inference latency. This is the default approach used by PyTorch.
They can also create multiple instances of the model being served, each with a single thread processing batches in parallel, with the hope of improving throughput. Another system, ParaX~\cite{parax}, does this to improve throughput.
Unfortunately, users are mainly left choosing between these two extremes;
while multi-instance execution will frequently maximize throughput at the expense of latency, only relying on intra-op parallelism frequently results in neither the best throughput nor the best latency.

In summary, our main observation is that while some intra-op parallelism improves inference throughput and latency across models and batch sizes, scaling such parallelism across more cores provides diminishing gains.
The key idea in \sys builds on this observation; stated crudely, \sys runs multiple instances concurrently and partitions all available threads in the system across these instances with the goal of minimizing inference latency for a given batch size.

In the next section, we formally state the problem we are trying to solve in \sys, and describe \sys{}'s systems design and algorithms that solve this problem.

\section{\sysname{} Design}
\label{sec:design}

\begin{figure}
\centering
    \includegraphics[width=1.0\columnwidth]{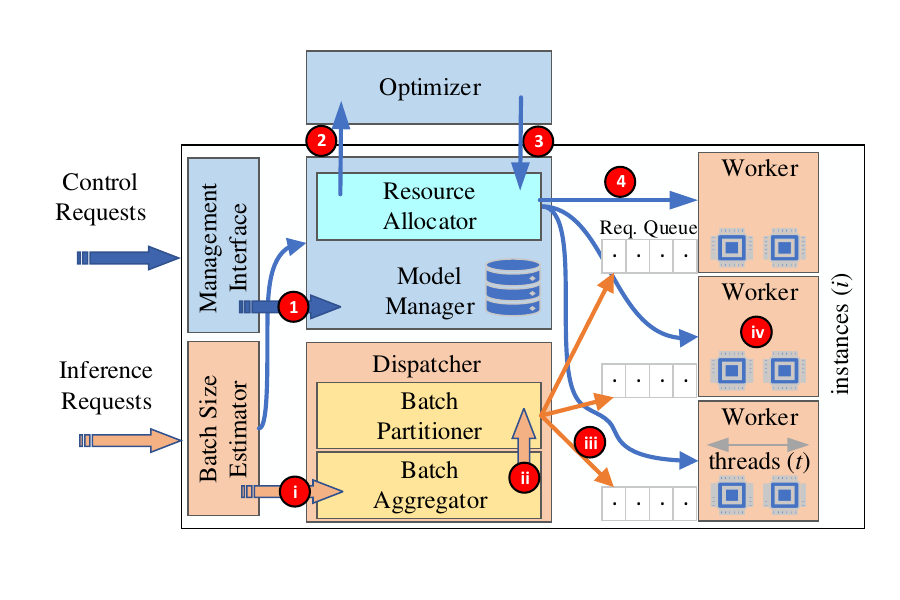}
    \caption{Architectural overview of \sys{}, highlighting its components and their interactions for the flow of inference requests (orange arrows) and for control messages such as configuration changes (blue arrows).}
    \label{fig:arch}
\end{figure}

We build \sys as an extension to TorchServe (a well-established serving system).
\sys is targeted at \emph{online} inference workloads, where input inference requests are constantly streaming in and responses need to be streamed out.
When \sys is enabled in TorchServe, it monitors incoming inference requests to select an appropriate batch size $B$, and transparently and dynamically reconfigures the number of model instances and the intra-op parallelism of each instance to improve average batch latency.
In cases where inference request rates change, this configured batch size might need to change as well, triggering re-configuration.

The key idea in \sys is that rather than having a single ``fat'' instance that uses all available threads to parallelize inference within a single batch, it instead divides large batches into smaller batches each processed concurrently by one of several ``thin'' instances that use a limited amount of intra-op parallelism.
Given a server with $T$ threads and incoming inference requests grouped into batches of size $B$, \sys determines a configuration $[\langle{}i_1$, $t_1$, $b_1\rangle{}, \ldots, \langle{}i_n$, $t_n$, $b_n\rangle{}]$ 
such that $\sum_{j=1}^{n} i_j \cdot t_j = T$ and $\sum_{j=1}^{n} i_j \cdot b_j = B$. For simplicity, we will refer to this list as a \emph{\itbitb{} configuration} for the remainder of this paper. In each $\langle{}i_j$, $t_j$, $b_j\rangle{}$ configuration in this list, $i_j$ instances concurrently execute model inference ($i_j$ specifies the number of instances of this type). 
Each such instance uses $t_j$ threads for intra-op parallelism ($t_j$ is the degree of intra-op parallelism for this instance), and the batch size processed by this instance is $b_j$.

For a given \itbTB{}, \sys tries to choose a configuration that minimizes average per-batch latency while improving throughput compared to using [\itb{1, $T$, $B$}] (the fat configuration).
However, determining the optimal \itbitb{} configuration is challenging because it is workload- and deployment-dependent: on the specific model being served, input dimensionality such as tokens in a sequence and batch size, and the number of cores and sockets in the targeted hardware. 
Furthermore, \sys{} has to determine the new optimal \itbitb{} configuration and reconfigure serving instances appropriately without stalling the entire serving system when the serving batch size changes (when request arrival rate changes), even though this might be infrequent (order of hours, not seconds)~\cite{fb1}.

To tackle these challenges, \sys{} uses a combination of workload profiling, algorithmic techniques to determine the optimal \itbitb{} configuration, and systems optimization to seamlessly reconfigure serving instances.
\sys{} profiles a range of single-instance configurations (\itb{$1$, $t$, $b$} configurations), then uses the measured average batch latency of each single-instance configuration to compute the \itbitb{} configuration that minimizes expected average batch latency by using a 2D dynamic-programming-based knapsack solution for a given model and \itbTB{}.
In some cases, an \itbitb{} configuration with only a single element is insufficient to describe the optimal configuration; \sys handles these cases, but we defer discussion of it to \S\ref{sec:non-uniform} to simplify exposition.
%

\sys uses a simple yet effective technique to reconfigure the serving system with an updated \itbitb{} configuration without any service downtime. 
It maintains two sets of instances, one active and one passive, and it reconfigures the passive set with the desired new configuration.
It then swaps the two sets, while scaling up the new active set while simultaneously scaling down the old active (now passive) set. 

\subsection{Architecture}

Figure~\ref{fig:arch} shows the overall architecture of the system.
A high-level flow of the system is as follows:

\begin{description}
    \item[Batch Size Estimator~(\S\ref{sec:estimation}).] The batch-size estimator
        estimates the number of requests that will arrive in a given time interval based
        on user requirements.

    \item[Optimizer~(\S\ref{sec:optimizer}).] If the estimator decides that the system needs to be reconfigured to serve a batch size $B$ in steady state, $B$ together with
        the number of cores~($T$) is fed to the optimizer to find the optimal
        \itbitb{} configuration for serving. The optimizer uses profiled
        data~(\S\ref{sec:profiling}) to find this optimal configuration.

    \item[Resource Allocator~(\S\ref{sec:resource_allocator}).] The resource
        allocator allocates resources to the instances based on the configuration
        found by the optimizer.

    \item[Dispatcher~(\S\ref{sec:dispatcher}).] Once the resources are
        allocated and all instances of the new optimal \itbitb{} configuration are created, the dispatcher forwards inputs to each instance as appropriate.

    \item[Worker~(\S\ref{sec:worker}).] Each instance executes the
        inference given to it by the dispatcher and then returns results.
\end{description}

We describe each of these components in detail next.

\subsection{Profiling}
\label{sec:profiling}

\sys uses model profiles to find \itbitb{} configurations that will improve performance for a given \itbTB{}.
Model profiling is always done using a single instance at a time, while varying threads for intra-op parallelism ($t$) and batch size ($b$).
The profiler runs configurations for various \itb{$t$, $b$} values.
In practice, we use \itb{$t$, $b$} $\in \{1, \ldots{}, T\} \times \{2^0, 2^1, \ldots{}, 2^{n}\}$.
For each of these configurations it records its average batch latency $L_{t, b}$.
By using only powers of 2 for $b$, \sys reduces the number of profiled configurations from $2^{n} \cdot T$ to $(n+1)\cdot T$.
Profiling more configurations could lead to more accurate performance estimates (and thus improve the optimizer's final choice of configurations), but \S\ref{sec:evaluation} shows this modest amount of profiling is sufficient to show substantial gains.
Moreover, profiling each configuration takes on the order of minutes, making profiling such a combinatorial space impractical for realistic workloads.
For example, for $n = 10$ and $T = 16$, using only powers of 2 for $b$ reduces the number of profiled configurations from 16,384 to 176
which reduces wall-clock profiling time from 30 days to \emph{a few hours}.
The profiler could also perform separate profiling to let \sys choose between options like ``eager'' versus ``graph'' mode; currently, \sys{}'s profiler uses whichever configuration options the user plans to run the model with.
Profiling is performed offline, and it is not on the inference critical path.

\begin{figure}
    \centering
    \includegraphics[width=1.0\linewidth]{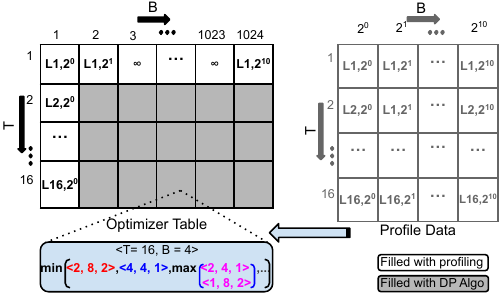}
    \caption{\sys{}'s latency minimization algorithm uses dynamnic programming. Single-instance profiled latency data is provided as input to the algorithm, which then fills in a comprehensive optimizer table for all values of \itbtb{}$\in$\itbTB{}.}
    \label{fig:optimizer}
\end{figure}

As Figure~\ref{fig:optimizer} shows, \sys{}'s optimizer queries the profiled lookup table to find the expected latency for a given configuration.
For each profiled configuration \itb{1, $t$, $b$}, \texttt{Profile$[\,t$,$b]\,$} contains the measured single-instance average batch latency (
represented as $L_{t,b}$), which the optimizer uses to find configurations to minimize end-to-end latency.

\subsection{Optimizer}
\label{sec:optimizer}

The optimizer is the core algorithmic component of \sys.
Its goal is to find an \itbitb{} configuration that minimizes average batch latency for a given \itbTB{}.
Optimal configurations for a given
\itbTB{} are cached to avoid repeated work.

\sys uses dynamic programming to find the expected optimal configuration for a given \itbTB{},
using the latency of the profiled configurations as an input.
We use a \textit{multi-dimensional knapsack} problem formulation~\cite{knapsack}.
The size of the knapsack is 2-dimensional;
the first dimension is the number of cores~($T$) and the second dimension is the
batch size~($B$).
Profiled configurations are used as the items to fill the knapsack. The
weight of each item is \itb{$t$, $b$}, and the value of the item is the expected average batch latency of
the \itbtb{} configuration. We can use a given \itbtb{} configuration multiple times (corresponds to the same \itbtb{} configuration executing concurrently).
The goal of the optimizer is to find a set of items that minimizes
average batch latency (\autoref{eq:knapsack1}) across model instances while keeping the total weight of the
items equal to the size of the knapsack~\itbTB{}~(\autoref{eq:knapsack2}).
\begin{align}
     & \text{Minimize} \max_{0 \le t_j \le T\atop 0 \le b_j \le B} L_{t_j, b_j} \label{eq:knapsack1} \\
     & \text{subject to} \sum{t_j} = T \mbox{and} \sum{b_j} = B \label{eq:knapsack2}
\end{align}

$L_{t_j, b_j}$ is the latency of the $\langle{}t_j$, $b_j\rangle{}$ configuration. $t_j, b_j$ are the number of cores and batch size of the $j$th configuration.

We can now describe our dynamic programming algorithm. Let $\mbox{opt}[t, b]$ be the total latency of processing $b$ inputs
with the $t$ threads. $\mbox{opt}[t, b]$ has the optimal sub-problem property: we can compute $\mbox{opt}[t, b]$ by looking at
$\mbox{opt}[t', b']$ where $t' \leq t$ and $b' \leq b$. If possible, $\mbox{opt}[t, b]$ is initialized to the
profiled latency with the same number of inputs and threads; otherwise, it is initialized to $\infty$.
Mathematically, $\mbox{opt}[t, b]$ can be computed as follows:
\begin{equation}
    \mathop{\mbox{opt}[t, b]} = \min \left(
        \max_{\substack{t' \leq t, b' \leq b}}
        \left(\mbox{opt}[t - t', b - b'], L_{t', b'}\right) \right)\nonumber
\end{equation}
where $L_{t', b'}$ is the latency of the profiled configuration
$\langle{}t'$, $b'\rangle{}$. The inner $\max$ is performed since the end-to-end latency of two concurrent work items is just the latency of the slower work item.
The returned configuration is then the one corresponding to $\mbox{opt}[T, B]$.
This algorithm has runtime complexity pseudo-polynomial in $T$ and $B$, which is practical for reasonable $T$ and $B$ values.

The above algorithm provides the optimal
solution in theory since it searches over all possible configurations.
However, in practice, the generated \itbTB{} solution might not match the expected theoretical optimal, since the optimizer depends on profiles measured in isolation, and it disregards performance contention from running various \itbitb{} configurations concurrently on the same multicore server (such contention profiling across all configuration combinations is impractical). We show in \S\ref{sec:speedup-gap} that the gap between the optimal solution in theory and practice is small.

\subsection{Resource Allocator}
\label{sec:resource_allocator}

The resource allocator assigns resources to instances based on the \itbitb{} configuration returned by the optimizer.
The resource allocator is the only component that interacts with the dispatcher
and the worker.
For now, the allocator assumes that resources are not over-subscribed and
$\sum i_j \cdot t_j$ is less than or equal to the number of physical cores in the system.
Given that the resources are not over-subscribed, the allocator can allocate
resources to the instances in a round-robin fashion.
The compute resources for each instance are statically allocated at the time of
instance creation and do not change at runtime. Hence, the allocator pins the
instance to the cores allocated to it to avoid thread migration
costs.

The allocator is independent of the optimizer and a user can specify other ways
to allocate resources to the instances. For example, the user can specify
specific cores or sockets for each instance.
By default, the allocator avoids assigning cores across
sockets to any single instance. This is done to avoid performance degradation due to
inter-socket communication overheads across NUMA domains.
However, while individual instances are socket-local, different instances can utilize all available
sockets in the system.

\subsection{Dispatcher}
\label{sec:dispatcher}

The dispatcher handles two types of requests: (1) management requests and (2) inference requests.
The dispatcher's management interface handles ``control'' messages such as requests to register a new model and those to create and delete
instances of any of the registered models. 
Management requests are handled in the dispatcher itself and are not on the critical path of inference execution.

Inference requests are dispatched to appropriate worker instances. 
The dispatcher itself handles both batch aggregation and  batch partitioning of the requests. 
Batch aggregation~($B$) is done per model and batch partitioning is done per instance using the $b$ values in the \itbitb{} configuration. 
Batch aggregation also uses a user-provided batch timeout value; request aggregation is done until the timeout expires.
If the timeout expires before the batch size~$B$ is reached, the dispatcher
simply dispatches the current batch to the instances.
The Batch Size Estimator triggers a configuration change if batch timeouts happen too frequently.
However, instance reconfiguration is time consuming and is done conservatively.

\subsection{Worker Instance}
\label{sec:worker}

Each worker instance is responsible for executing an inference batch with~$b$ inputs for a
given model using $t$ threads. Each worker executes a user-provided handler over
a batch of requests. A handler takes the batch of requests as input and returns
the batch of responses.
During the handler initialization, the worker might need to load the model into the memory. 
Users may also specify any optimizations to use during model initialization. 
For example, the user can specify that the model should be loaded and optimized for graph mode~(TorchScript for PyTorch framework).
Each handler mainly consists of three parts: (1) Pre-processing, (2) Inference and (3) Post-processing.
Pre-processing and post-processing are user provided functions which are executed before and after inference. 
Inference is executed by the framework (e.g., PyTorch).
Pre-processing usually involves data transformations. For example, 
pre-processing for an image classification model can involve decoding the image, resizing it
to a fixed size, and transforming it into the right format for the model (e.g., a PyTorch tensor).
Post-processing usually converts the output of the inference into user-understandable format. For example, for a computer vision model, 
post-processing can be the conversion of the output tensor to a list of
labels.

Inference is executed by the framework using parallel implementations of the operators~(intra-op parallelism). 
Each parallel operator implementation is responsible for executing the operator across $t$ intra-op threads.
%
This parallelization involves slicing the input batch into multiple chunks, partitioning operator state across threads, and executing the operator on each chunk in parallel.
\sys does not improve the mechanism of operator parallelization but simply uses the functionality provided by the framework in a more efficient way by assigning an appropriate number of threads.

\subsection{Configuration Changes}
\label{sec:configuration_changes}
Reconfiguration is the process of changing the \itbitb{} configuration for a model and is handled by \sys{}'s Resource Allocator.
The Batch Size Estimator triggers a configuration change by invoking the optimizer with a new batch size $\widetilde{B}$ if it predicts that the request arrival rate for a given model has changed considerably.
Reconfiguration does not generally require \sys{} to run any new profiling; as the batch size changes, the optimizer is re-run with the new $\widetilde{B}$ value to find the right configuration for the new batch size (if the given \itbTB{} configuration is not present in the optimizer cache).

Reconfiguration is time consuming and done conservatively, i.e., the configuration change is only initiated if batch aggregation timeouts are being triggered frequently or if request queuing delays are large, and this is ongoing for an extended time period (\S\ref{sec:estimation}). 
\sys works with an implicit assumption that the workload for a given model does not change frequently, which is a reasonable
assumption for many datacenter workloads~\cite{fb1}.
Moreover, dramatic workload changes would not only affect \sys configuration but could also require datacenter-level resource reprovisioning.

\sys uses a \torchserve feature, worker scaling, to handle configuration changes.
Worker scaling is the process of increasing or decreasing the number of workers for a given model.
However, in \sys, we might have to change the configuration of the model instance itself, as dictated by the optimizer, by allocating it fewer or more threads than currently assigned.

The current \sys implementation handles the configuration change in two different ways.
The first is when a configuration change only requires increase or decrease in the
number of instances, but the number of threads within each of the existing instances remains the same. 
Such configuration changes are handled by the worker scaling mechanism.
Scaling down is achieved by removing the workers of a model one by one. 
Workers are removed in a round-robin fashion and resources are released back to
the resource allocator~(\S\ref{sec:resource_allocator}).
Scaling up is similar to the initial worker creation process. 
%

The second is the trickier case, and it occurs when the configuration change requires different number of threads for
the workers as compared to their current configuration.  
\sys handles such reconfigurations in a two-step process called active-passive scaling~(\S\ref{sec:active_passive_scaling}).
\sys relies on this two-step process to avoid changing the internal operator
implementation libraries and making our approach portable across serving
systems. Operator implementation libraries like ATen, MKL-DNN, etc.
have their own internal mechanisms to manage and schedule the
threads~\cite{openmp-docs}, but these libraries are not designed to handle 
frequent configuration changes.
For example, PyTorch allows changing the number of MKL
threads for each instance when the library is used with
\texttt{MKL\_DYNAMIC=true};  however, to implement this MKL creates and
destroys threads for each matrix multiplication, resulting in lower
performance due to the high cost of creating threads~\cite{mkl-issue}. Hence, PyTorch uses internal MKL libraries with
\texttt{MKL\_DYNAMIC=false}.

\subsubsection{Active-Passive Scaling.}
\label{sec:active_passive_scaling}

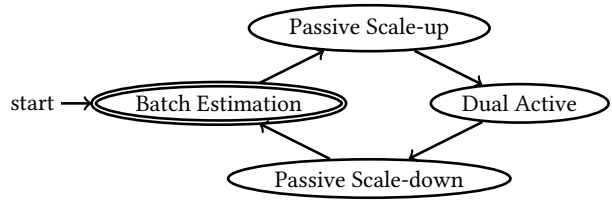
\begin{figure}
    \centering
    \begin{tikzpicture}[line width=1pt, inner sep=2pt]
        \node (estimation) at (0,0) [ellipse, initial, accepting, draw] {Batch Estimation};
        \node (reconfig) at (2,1) [ellipse, draw] {Passive Scale-up};
        \node (dual) at (4,0) [ellipse, draw] {Dual Active};
        \node (finish) at (2,-1) [ellipse, draw] {Passive Scale-down};

        \draw[->] (estimation) -- (reconfig);
        \draw[->] (reconfig) -- (dual);
        \draw[->] (dual) -- (finish);
        \draw[->] (finish) -- (estimation);
    \end{tikzpicture}
    \caption{State transitions for Active-Passive Scaling.}
    \label{fig:active-passive-scaling}
\end{figure}

\sys uses active-passive scaling when the optimizer's suggested configuration change requires instances to adjust the number of threads allocated to instances.
A naive way of going about such a reconfiguration would be to first shut down all instances in the old configuration (e.g, \itb{$i_{1}$, $t_{1}$, $b_{1}$}) and then start all instances in the new configuration (e.g., \itb{$i_{2}$, $t_{2}$, $b_{2}$}).  
In the worst case, all the old workers will be removed, and new workers will be created.
However, such an approach risks having the serving system be unresponsive for the entire duration of such time-consuming reconfigurations.

\sys uses active-passive scaling to avoid disruption.
For each model, \sys maintains two versions of the model. 
The active version respects the current configuration and is currently serving requests. 
The passive version has zero workers and stays inactive until activated. 

Active-passive scaling
is done in three steps (Figure~\ref{fig:active-passive-scaling}).
First, the passive version is scaled up to the new
configuration (e.g., we scale up to $i_2$ workers as per the new \itb{$i_{2}$,
$t_{2}$, $b_{2}$} configuration).  Next, the dispatcher starts redirecting new
requests to the new passive instances.  Finally, the historically
active version is scaled down to zero workers in the background (from $i_1$
workers as per the old \itb{$i_{1}$, $t_{1}$, $b_{1}$} configuration) once they
have completed their ongoing requests and been deactivated at the dispatcher.
At this point, the active and passive sets of workers have been swapped.

\subsection{Batch Size Estimation}
\label{sec:estimation}

To choose a good configuration, \sys{} needs to know the batch size for the
current workload ($B$). \sys{} estimates the batch size in an online fashion by
tracking the request queue depth over time. It is easy enough for the Batch
Aggregator to track the size of each batch that it passes to workers, but
this batch size varies over time depending on input request arrivals, and different
batch sizes have different ``optimal'' \itbitb{} configurations.

Reconfiguring the number of instances and threads takes
several seconds and is expensive (\S\ref{sec:config-change-latency}), so it is
important that reconfiguration only happens when the workload is stable enough
to warrant it.  Without some kind of smoothing, \sys{} will risk 
``flip-flopping'' between configurations. Packrat uses two-level smoothing to
avoid this problem.
First, Packrat's Batch Size Estimator uses the most recent request queue depth
$\widehat{Q}$ to track an exponentially weighted moving average of request queue
depth ($\widetilde{Q}_x = \alpha{}\widehat{Q} + (1 - \alpha)\widetilde{Q}_{x-1}$)
and picks the next lower power of two to $\widetilde{Q}$ as an estimated batch size
$\widehat{B}_x$. Second, Batch Size Estimator takes the mode over the last $n$
estimated batch sizes ($\widehat{B}_{x-n}, \ldots{}, \widehat{B}_x$) to get a final smoothed batch size
($\widetilde{B}$).
After each reconfiguration timeout, Packrat's Batch Size Estimator compares the
current batch size $B$ to the smoothed batch size $\widetilde{B}$. If
$\widetilde{B}$ is different from $B$, Packrat reconfigures the
system to use the new batch size $\widetilde{B}$.
\S\ref{sec:config-change-latency} shows that
this simple approach works well in practice, and \sys{} uses it to both scale up
and scale down the batch size $B$ as request arrival rates change.


\section{Implementation}
\label{sec:implementation}
We implement \sys as an extension to TorchServe~\cite{torchserve}, a popular serving system in the PyTorch ecosystem. 
We use PyTorch for all benchmarking results in this paper and TorchServe for all end-to-end experiments.
We use TorchServe as the base serving system for \sys because it is easy to use
and provides a simple REST API for serving models. TorchServe also provides features for model management,
adaptive batching, a management API for worker creation and deletion, and a simple
REST API for serving models.

\sys augments TorchServe with features such as the batch size estimator, batch aggregator, optimizer, and
resource allocator.
TorchServe supports plugins to customize the serving system.  For example, we can customize the batching layer to provide our custom
batch aggregation, batch partitioning, and batch size estimation strategies, and
the worker management layer for our custom resource allocator.
In all, \sys is implemented in $\sim$~5k lines of code. This code will be made public with the final version of the paper.

\paragraph{Optimizer.} The optimizer is responsible for providing the optimal
configuration for a given \itbTB{} pair. However, the optimizer does not
directly interact with the resource allocator.
We implement the optimizer as a standalone service.
A separate task acts as a client to the optimizer and uses TorchServe's
management API to communicate with the resource allocator.
The resource allocator then updates the configuration to match the desired configuration returned by the optimizer.

\paragraph{Resource allocation.} \torchserve supports custom resource allocators that can create and destroy workers or can allocate resources to workers.
For fault tolerance, \torchserve also respawns workers if they die.
\sys's custom \torchserve resource allocator maintains information about all idle and busy cores and the desired \sys configuration.
Based on the target configuration, it creates workers on demand, and it destroys them when they
are not needed. Internally, it uses a modified version of Intel's IPEX
\torchserve launcher to launch and pin workers threads on the desired cores~\cite{launcher}.
This custom resource allocator is only used when \sys is enabled in \torchserve.

\paragraph{Batch aggregation and request estimation.} \torchserve allows custom batching algorithms that can be integrated into the system at startup time. 
Our batch aggregator extends TorchServe's default implementation; we add a batch size estimator which intercepts incoming requests
and estimates the batch size for each inference endpoint~(\S\ref{sec:estimation}).


\section{Evaluation}
\label{sec:evaluation}

We evaluate \sys{} using both inference microbenchmarks with PyTorch and end-to-end performance with \torchserve.  Our evaluation seeks to answer the following questions:
\begin{itemize}
    \item How does \sys{}'s proposed approach compare to the state-of-the-art for inference microbenchmarks with PyTorch for a range of models and batch sizes? (\S\ref{sec:ubench-speedup})
    \item How accurate is the optimizer in predicting the performance of multi-instance configurations? (\S\ref{sec:speedup-gap})
    \item Is \sys{}'s optimizer general enough to generate configurations for deployments that have non-powers-of-two threads per socket? (\S\ref{sec:non-uniform})
    \item How does \sys{} improve end-to-end serving latency and throughput in \torchserve, and how do these numbers compare to microbenchmarks with PyTorch? (\S\ref{sec:eval-single-model})
    \item How effective is \sys{}'s reconfiguration in avoiding stalls and improving serving latency? (\S\ref{sec:config-change-latency})
\end{itemize}

\subsection{Experimental Setup}
\label{sec:experimental-setup}

\begin{table}[t]
    \centering
    \begin{tabular}{p{0.15\columnwidth} p{0.75\columnwidth}}
        \toprule
        \textbf{CPU}      & $2\times$ 16-core Intel Xeon Gold 6142 at 2.6 GHz
        \\
        \textbf{RAM}      & 384GB~(6x32 GB DDR4-2666 DIMMs per Socket)
        \\
        \textbf{OS}       & Ubuntu 20.04 LTS, Linux 5.4.0-100-generic
        \\
        \textbf{Software} & Python 3.8.10, PyTorch 1.12.1, TorchServe 0.6.1, Intel
        MKL-DNN v2.6.0,  OpenMP 4.5
        \\
        \bottomrule
    \end{tabular}
    \caption{Server configuration for all our experiments.}
    \label{table:setup}
\end{table}

All experiments are performed on a single Cloudlab~\cite{cloudlab}
machine~(c6420).
The hardware and software configuration of the machine is shown in
Table~\ref{table:setup}.
Hyperthreading is disabled on the machine for all experiments since
inference workloads extensively use the fused-and-multiple~(FMA) units shared
between hyperthreads. OpenMP threads scheduled on the same core will compete for
the FMA units, and performance is affected if hyperthreading is enabled~\cite{hyperthreading}.
Each benchmark is run for 100 iterations and the average time is reported. We use
a warmup phase of 10 iterations for each benchmark to ensure that
execution is not affected by startup overheads.

\subsection{Microbenchmarks}
\label{sec:microbenchmarks}

\begin{figure*}[t]
    \centering
    \includegraphics[width = 1.0\textwidth]{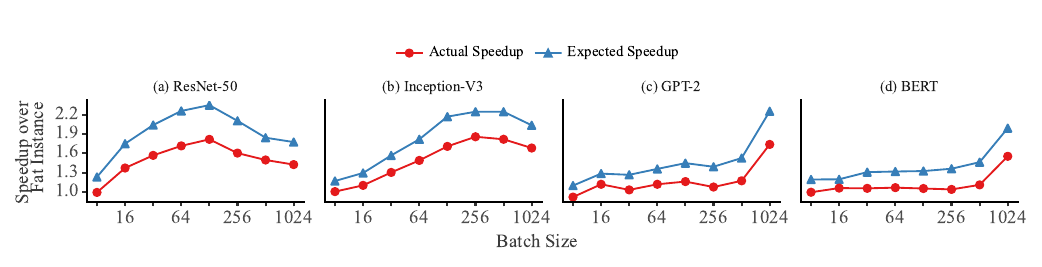}
    \caption{Inference microbenchmark when using \sys{} in PyTorch, showing the \textbf{speedup} of \sys over baseline fat-instance execution in graph mode for four different DNN models.  We also see the comparison of \sys{}'s expected speedup (estimated from isolated runs of individual instances) to the actual speedup attained when running multiple thin instances concurrently.}
    \label{fig:speedup-latency}
\end{figure*}

\paragraph*{Setup.} Microbenchmarking is performed across various models and
batch sizes. All microbenchmarks are written using PyTorch and use
pre-trained models downloaded from the PyTorch model zoo. Each model is
evaluated both in eager and graph modes.
In the microbenchmarks, we report the time taken by the model forward pass when using PyTorch (with and without \sys).
Later in \S\ref{sec:eval-sys}, we report the end-to-end latency of inference requests when using TorchServe (with and without \sys).
Unless explicitly stated, we report \sys speedup numbers compared to corresponding fat-instance baselines.
Similarly, unless explicitly called out, we only report results using graph mode as it consistently outperforms eager mode by providing lower latency in all our experiments, and thus is a stronger baseline.

\paragraph*{Models.} We microbenchmark the following models:
ResNet-50~\cite{resnet}, Inception-v3~\cite{inception}, GPT-2~\cite{gpt2}, and
BERT~\cite{bert}.
The ResNet-50 and Inception-v3 models are image classification models, GPT-2 is a text generation model, and
BERT is a text classification model.
These models are popular and widely used in real applications.

\subsubsection{Speedup over Baseline Execution.}
\label{sec:ubench-speedup}

Figure~\ref{fig:speedup-latency} shows the
throughput and latency speedup of multi-instance execution over
fat instances for ResNet-50~(a), Inception-v3~(b), GPT-2~(c), and
BERT~(d) models.
The speedup is measured for different batch sizes and for all threads in a socket. 
The fat instance is run with 16 threads and batch size $B$ and the thin instances use the \itbitb{} configuration suggested by \sys{}'s optimizer where \itbTB{} is partitioned across $\sum i_j$ smaller instances where $\sum i_j \cdot t_j = T$ and $\sum i_j \cdot b_j = B$.
For a given \itbTB{}, we measure the average throughput and latency of \sys{}'s chosen configuration ($\tau_P$ and $\lambda_P$) and of the fat-instance baseline ($\tau_B$ and $\lambda_B$). Throughput and latency speedups are calculated as $\tau_P/\tau_B$ and $\lambda_B/\lambda_P$, respectively. In practice, our measured throughput and latency speedup are almost always the same.

Even though \sys{}'s chosen configurations use the same total number of threads as the fat instance, \sys obtains substantial improvements in latency and throughput. The image classifiers, ResNet50 and Inception-V3 show a 1.53$\times$ and 1.52$\times$ mean speedup across batch sizes, respectively; the language models GPT-2 and BERT show a 1.18$\times$ and 1.13$\times$ average speedup, respectively.

There are two key reasons that \sys{}'s configurations outperform the fat instance which uses all threads on the server for intra-op parallelism.
First, all OpenMP threads synchronize at multiple barriers in the
fat-instance execution resulting in compute resource under-utilization.
However, in multi-instance execution, thread(s) in each instance can execute
independently of other instances. This allows the multi-instance execution
to utilize the available compute resources more efficiently.

Second, usually, workloads have multiple phases with different characteristics (e.g., a part that is compute-intensive and another that is memory-intensive). OpenMP barrier sync enforces all the threads
to march in lock-step, forcing every thread to execute similar work. This results
in over-utilization of one resource and under-utilization of other resources.  However,
\sys{}'s configurations include some degree of multi-instance execution; hence, the threads in each instance can execute
different phases without coordination. This results in better average compute and memory bandwidth utilization, which is also apparent when profiling the execution of both approaches.

As noted in \S\ref{sec:profiling}, \sys's profiler does not exhaustively profile all possible \itb{$t$, $b$} configurations to keep total profiling time on order of hours (instead of days). We found empirically that exhaustive profiling (all values of $b$) does not change the \sys-selected configurations and hence actual latency speedups.

\paragraph*{Why not just use many single-threaded instances?}
\label{sec:thin-comparison}

\begin{figure}[t]
    \centering
    \includegraphics{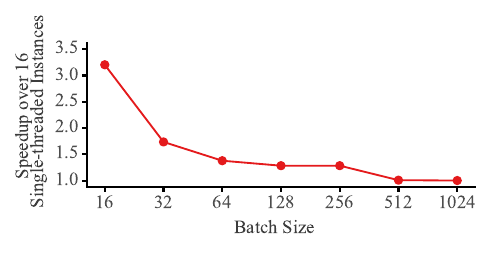}
    \caption{Speedup of \sys{} when compared with 16 single-threaded instances for GPT-2. \sys{} always exceeds or matches 16 single-threaded instance performance.}
    \label{fig:thin-comparison}
\end{figure}

Fat instances are one baseline, but another easy-to-configure baseline is to run $T$ single-threaded instances, one on each CPU core.
Single-threaded instances sometimes help; it frequently performs worse than the fat instance baseline.
In contrast, \sys{} always provides some speedup against the fat instance baseline (Figure~\ref{fig:speedup-latency}), and it always exceeds or matches $T$ single-threaded instances~(as shown in Figure~\ref{fig:thin-comparison}).
For other models, \sys{} performs either better (ranging from 1.02$\times$ to 1.75$\times$) or equal to 16 single-threaded instances.
We can speculate on why single-threaded instances perform poorly. In our experience, $T$ single-threaded instances suffer from more cross-instance resource interference and under-utilization (see \S{\ref{sec:speedup-gap}). Single-threaded instances also inherently cannot exploit intra-op parallelism to improve latency. Regardless, \sys{} finds high-performing configurations.

\subsubsection{Expected versus Observed Speedup.}
\label{sec:speedup-gap}

\begin{figure}[t]
    \centering
    \includegraphics{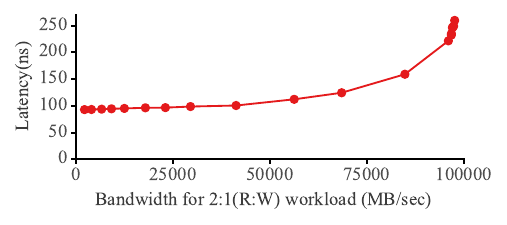}
    \caption{Effect of memory bandwidth load on access latency.}
    \label{fig:memory-lat}
\end{figure}

Figure~\ref{fig:speedup-latency} shows that while \sys{} is able to obtain significant latency benefits over the baseline, the attained speedup for all models (across batch sizes) is less than \sys{}'s expected speedup \--- this can be seen in the gap between the \texttt{Expected Speedup} and \texttt{Actual Speedup} lines.
This is because \sys{} estimates the performance of configurations composed of multiple instances running concurrently based on profiling of individual instances run in isolation; however, the actual performance of each instance when run concurrently with other instances is lower since concurrent execution creates some resource contention between the instances.
For example, running \sys{}'s configuration in practice creates more CPU and memory bandwidth interference compared to when profiling was done.
Next, we dive deeper to understand and account for these sources of contention.

\paragraph*{License-based downclocking.}
The first major source of interference between instances is due to
license-based
CPU downclocking~\cite{downclocking}. License-based downclocking is a mechanism 
used by CPU vendors to limit CPU frequency when many cores use SIMD
instructions concurrently for sustained periods. This is done for energy efficiency
reasons~\cite{downclocking}.
For example, even though the normal CPU frequency for an Intel Xeon Gold 6142 is
2.6~GHz~(Table~\ref{table:setup}), its frequency is downclocked to 2.2~GHz when 
all cores run SIMD instructions concurrently~\cite{wikichip}. This
lowers each core's performance by about 15\%; we experimentally show this explains about half of the gap between expected and observed speedups below.

\paragraph*{Loaded memory latency.}
The second source of interference is due to increased load on the memory controller.
Ideally, cores would be able to use any available memory bandwidth without impacting other cores so long as memory bandwidth isn't saturated; however, in practice, memory bandwidth load created by one instance increases effective memory access latency for other instances.
Figure~\ref{fig:memory-lat} shows this effect by measuring memory access latency under varying memory bandwidth load;
this microbenchmark uses a 2:1 read-write ratio similar to our inference workloads. 
The increased memory latency explains the other half of the gap
between expected and observed speedups.

\paragraph*{In-depth analysis for ResNet-50 microbenchmark.}
To verify that downclocking and degraded memory access latency explain the difference between expected and actual performance,
we perform an
in-depth analysis using the ResNet-50 microbenchmark.
As shown in Figure~\ref{fig:speedup-latency}~(a), due to the overheads of the multi-instance execution the gap between expected and realized speedups is between 12-15\%.
To show the impact of license-based downclocking on inference performance, we implement a SIMD load generator that saturates the FMA units on a configurable number of cores.
We monitor the performance of a single thin instance while we run the SIMD load generator on the cores that are not being used for inference (\autoref{fig:gap-analysis}, \texttt{Thin(1) + FPGen}).

Similarly, to show the effect of increased memory latency on inference performance, we implement a custom
load generator that generates a configurable amount of memory bandwidth load.
We use
this to generate load that is about equal to the load generated by $i - 1$
thin instances, which simulates the memory load of running a thin instance concurrently with $i - 1$ other instances~(\autoref{fig:gap-analysis}, \texttt{Thin(1) + MemGen}).

\begin{figure}
    \includegraphics[trim={0 0 0 0.2cm},clip,width = 1.0\linewidth]{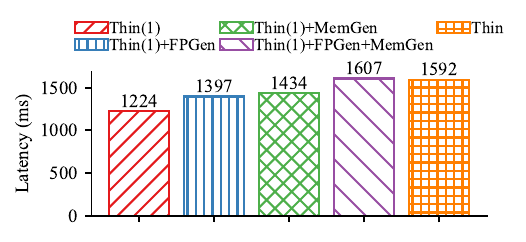}
    \caption{Breakdown of why multi-instance performance doesn't match the performance by predicted from profiling single-instance latency (\texttt{Thin(1)}). When combined with a SIMD (\texttt{Thin(1)+FPGen}) or a memory bandwidth load generator (\texttt{Thin(1)+MemGen}) or both (\texttt{Thin(1)+FPGen+MemGen}) a single instance's performances slows to match that measured when 16 instances run together (\texttt{Thin}).}
    \label{fig:gap-analysis}
\end{figure}

Figure~\ref{fig:gap-analysis} shows this analysis for a single configuration
($T = 16$, $B = 256$).  The optimizer recommends using 16 thin instances each with
($t = 1$, $b = 16$). The latency for the baseline fat instance is \ms{2664} and for a single
($t = 1$, $b = 16$) instance is \ms{1224} which is about a 54\% reduction over the baseline.
However, when we use 16 thin instances, the latency is \ms{1600} which is about
a 40\% reduction over the baseline fat instance. So, the actual latency
reduction is 14\% lower than the expected reduction. The goal of this analysis
is to understand the gap between single thin instance~(\texttt{Thin(1)}) and
multiple thin instances~(\texttt{Thin}) as shown in
Figure~\ref{fig:gap-analysis}.
The impact of license-based downclocking increase the latency of a single
thin instance to \ms{1397}~(\texttt{Thin(1) + FPGen}) and the impact of increased memory
latency degrades the latency of \texttt{Thin(1)} to \ms{1434}~(\texttt{Thin(1) + MemGen});
these are \ms{173} and \ms{210} higher (worse) than the isolated single thin-instance latency.
As reference points in Figure~\ref{fig:memory-lat},
\texttt{Thin(1)} generates memory traffic of around 3~GB/s and \texttt{Thin}
generates around 50~GB/s.
If we add all three overheads, we get around \ms{1600} which is the latency of multiple
thin instances latency~(\texttt{Thin}).
Hence, the combination of license-based downclocking and increased memory access latency explains 
the discrepancy between the expected latency estimated from profiling thin instances in isolation and the actual latency in practice.

\paragraph*{Why not model resource interference in the optimizer?}
\sys's Optimizer does not model the interference described in~\S\ref{sec:speedup-gap}.  While designing \sys, we hypothesized that since this interference affects all configurations in a similar way, modeling interference wouldn't change the selected configurations.

To validate this hypothesis, we did two things. For select models, we looked at the gap between actual and expected speedup (e.g., in Figure~\ref{fig:speedup-latency}). In all cases, actual is a constant factor slower than expected (reassuringly, the \emph{same} constant factor across models). Next for these models, we reran the Optimizer's DP taking into account the estimated interference performance penalty. The resulting configurations (and actual performance) was identical to the configurations selected when not accounting for interference. The reason is that if all profiled performance measurements are penalized by multiplying them by some constant $c < 1$, then their relative order doesn't change. The equation for $\mbox{opt}[t, b]$ still makes the same choices at each step even if all costs are multiplied by the same $c$.  \sys could be extended to incorporate modeled interference, but we have yet to find a case where doing so changes the chosen configuration.

\subsubsection{Non-Uniform Instances in \sys{}.}
\label{sec:non-uniform}

\begin{table}
    \centering
    \begin{tabular}{ccc}
        \toprule
        \textbf{Batch Size ($B$)} & \textbf{Cores ($T=16$)} & \textbf{Cores ($T=14$)}            \\ \midrule
        8                       & \itb{2, 8, 4}         & \itb{2, 7, 4}                    \\
        16                      & \itb{4, 4, 4}         & \itb{1, 6, 8}, \itb{2, 4, 4}     \\
        32                      & \itb{4, 4, 8}         & \itb{1, 6, 16}, \itb{2, 4, 8}    \\
        64                      & \itb{4, 4, 16}        & \itb{1, 6, 32}, \itb{2, 4, 16}   \\
        128                     & \itb{4, 4, 32}        & \itb{2, 7, 64}                   \\
        256                     & \itb{8, 2, 32}        & \itb{2, 3, 64}, \itb{4, 2, 32}   \\
        512                     & \itb{8, 2, 64}        & \itb{2, 3, 128}, \itb{4, 2, 64}  \\
        1024                    & \itb{8, 2, 128}       & \itb{2, 3, 256}, \itb{4, 2, 128} \\ \bottomrule
    \end{tabular}
    \caption{The best \itbitb{} configurations identified by \sys{}'s Optimizer for BERT with different batch sizes for two deployments with $T = 16$ and $T = 14$ respectively.}
    \label{table:non-uniform-bert}
\end{table}

For most of the benchmarks, the optimizer generates a uniform \itbitb{} configuration where each
thin instance has the same number of cores and the same batch size. This is because the number of threads ($T$) and batch size ($B$) used in most benchmarks are powers of two. In real-world scenarios, number of cores and batch sizes may not always be powers of two; hence,
we investigate the performance in these cases here. We show
the impact of such configurations on the inference latency and throughput. To conserve space, here we show the results
only for the BERT model. Results for other models are similar.
Table~\ref{table:non-uniform-bert} shows the \itbitb{} configurations for different batch sizes for
T = 16 and T = 14. For example, for $B = 16$, the configuration is ($i = 4$, $t = 4$, $b = 4$) for
T = 16; it is ($i = 1$, $t = 6$, $b = 8$) and ($i = 2$, $t = 4$, $b = 4$) for T = 14. So, the final
configuration includes a mix of different thin instance types. Similar configurations
are generated for other batch sizes.
For such cases, \sys{}'s optimizer chooses configurations where the latency of different instances types are similar, resulting in lower overall latency while satisfying Equations~\ref{eq:knapsack1} and \ref{eq:knapsack2}.

\subsection{End-to-End Experiments}
\label{sec:eval-sys}

Next, we evaluate \sys{}'s end-to-end performance for different models and 
batch sizes in the context of \torchserve. The underlying setup is the same as in \S\ref{sec:microbenchmarks}.
\sys is built as an extension to \torchserve~\cite{torchserve}; unless otherwise stated, we use the default configuration of \torchserve. 
We evaluated all models in both eager and graph modes.
\sys{}'s improvements are similar for both eager and graph execution mode, so we only show the results for graph mode execution since it has better baseline performance. 
%

\subsubsection{Latency Improvements.}
\label{sec:eval-single-model}

\begin{figure*}[t]
    \centering
    \includegraphics[width = 1.0\textwidth]{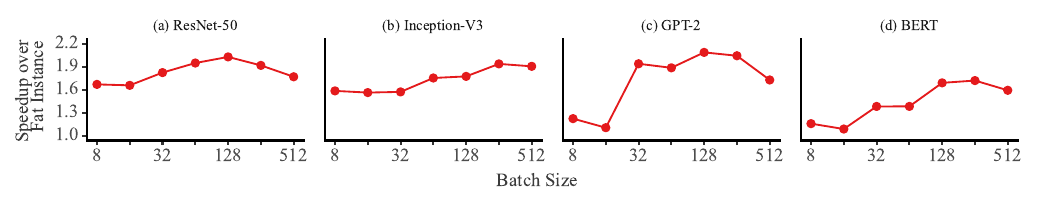}
    \caption{\sys{}'s end-to-end latency and throughput speedup over the corresponding baseline (fat-instance) runs in TorchServe. GPT-2 is the only model that uses eager mode in this figure; it crashes \torchserve in graph mode.}
    \label{fig:ts-speedup}
\end{figure*}

Figure~\ref{fig:ts-speedup} shows the latency and throughput speedup of \sys{}'s chosen configurations
over baseline fat-instance execution for ResNet-50~(a),
Inception-v3~(b), GPT-2~(c), and BERT~(d).
%
Table~\ref{table:mean-max-speedup} summarizes the speedup across all 
batch sizes for each model. \sys{} consistently improves performance
across all batch sizes for all models. \sys{} provides an average speedup of 1.43~to~1.83$\times$ and a maximum speedup of 1.72~to~2.09$\times$.

\noindent
\paragraph*{Gains compared to microbenchmarks.} \sys{}'s end-to-end gains on TorchServe are higher than in the microbenchmarks. This is because
the microbenchmarks only perform inference; with \sys on TorchServe,
we measure end-to-end impact on both inference and TorchServe's serving components as well.
For each inference batch, a worker executes a handler that consists of
pre-processing, inference, and post-processing. Since the pre-and post-processing
are usually not compute intensive, serving systems
typically use a single thread for them. However, multi-instance execution
parallelizes pre- and post-processing as well, resulting
in higher performance gains compared to microbenchmarks.

\begin{table}[t]
\centering
\begin{tabular}{lcccccc}
\toprule
    &
    \multicolumn{2}{c}{\textbf{PyTorch}} &
    \multicolumn{2}{c}{\textbf{PyTorch}} &
    \multicolumn{2}{c}{\textbf{TorchServe}} \\
    &
    \multicolumn{2}{c}{\textbf{[Eager]}} &
    \multicolumn{2}{c}{\textbf{[Graph]}} &
    \multicolumn{2}{c}{\textbf{[Graph]}} \\
    \textbf{Model} &
    \textbf{avg} & max &
    \textbf{avg} & max &
    \textbf{avg} & max \\ \midrule
ResNet50
    & \textbf{1.89} & 2.67
    & \textbf{1.53} & 1.83
    & \textbf{1.83} & 2.03 \\
Inception-V3
    & \textbf{1.53} & 2.11
    & \textbf{1.52} & 1.88
    & \textbf{1.73} & 1.94 \\
GPT-2
    & \textbf{1.19} & 1.74
    & \textbf{1.18} & 1.75
    & \textbf{1.71}\textsuperscript{*} & 2.09\textsuperscript{*} \\
BERT
    & \textbf{1.14} & 1.54
    & \textbf{1.13} & 1.57
    & \textbf{1.43} & 1.72 \\ \bottomrule
\end{tabular}
\caption{\emph{\textbf{\sys{}'s mean and max speedup}} across batch sizes in PyTorch (eager and graph mode) and \torchserve.
\textsuperscript{*}GPT-2 runs in eager mode because it crashes \torchserve in graph mode.}
\label{table:mean-max-speedup}
\end{table}

\subsubsection{Configuration Change Latency.}
\label{sec:config-change-latency}

\begin{figure}[t]
    \centering
    \includegraphics{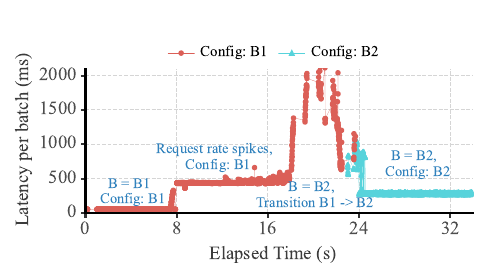}
    \caption{Configuration change in \sys{}. A spike in request rate at 8~s results in sub-optimal latencies under configuration $B1$.  Adjusting the batch size from $B1$ to $B2$ triggers a configuration change.  Once reconfigured to the new optimal \itbitb{}, \sys{}'s serving latency improves over the latency observed when it was using a sub-optimal configuration right after the request spike.}
    \label{fig:config-change}
\end{figure}

Finally, we evaluate the impact of configuration change using the Inception-v3 model, with input arrival rates following a step function. All of the cores on a single CPU socket are used for this experiment ($T = 16$).
Figure~\ref{fig:config-change} shows a zoomed-in timeline of model latency just before, during, and just after a configuration change.
The experiment starts with the multi-instance configuration for $B = 8$, which correctly corresponds to the load generated by the client and produces the expected batch size.
After some time, the request rate spikes, changing the ideal batch size for the workload to $B = 64$; 
however, we force the server to not activate a change in batch size immediately (to observe the performance impact of doing this) and the server continues to handle these batches with the $B = 8$ optimized configuration for some time.
Finally, the server starts the configuration change for $B = 64$ and begins handling requests with the new configuration.

There are five key takeaways in \autoref{fig:config-change}: (1) Response latencies are
initially stable.
(2) After 8s, the client increases the rate of input requests, and the server
handles requests with the old configuration until 18s. The average latency
increases significantly due to queuing delays.
(3) After 18s, the server starts the configuration change for $B = 64$.  The
average latency per batch increases due to the configuration change overhead
until 23s; the serving system does not stall processing requests during
reconfiguration.  The configuration change takes around 5s.  Most of this comes
from underlying systems~(confirmed with a microbenchmark on \torchserve).
Given such reconfigurations occur infrequently, we
consider this overhead reasonable.
(4) During reconfiguration (18-23s), the server handles requests
with the new configuration.  The average latency per batch jumps by
2-3$\times$ due to initialization overhead of the new configuration
and resource oversubscription, as both old and new configurations are active.
(5) After initialization completes, the server handles all requests with
the new configuration, and the average latency re-stabilizes. The new configuration reduces latency by 1.54$\times$ over the old configuration when $B=64$.


\section{Related Work}
\label{sec:relatedwork}

In this section, we describe relevant related work, and draw out differences to \sys{}.

\paragraph{Adaptive batching.}
Many recent works optimize batched network request processing~\cite{ix,shinjuku,zygos,shenango} or packet processing~\cite{batchy, preliminary}
on multi-core CPUs.
A common technique in these frameworks is to process a batch of requests to completion up to some batch size $B$. When processing completes another batch is collected similar to \sys{}. These systems face a similar problem in how they divide work across cores; for example, MICA~\cite{mica} showed that in some cases it makes sense to strictly partition state and work across cores, while with other workloads it can make more sense to balance request processing across cores and letting them all synchronize access to shared state. \sys{}'s techniques may be useful for navigating that trade-off automatically as well.

Adaptive batching has also been used to more efficiently serve classic ML and DNN models. Clipper uses adaptive batching to maximize
throughput subject to a compute processing latency~\cite{crankshaw2017clipper}.
Nexus tries to schedule the inference computations of
multiple models onto a given set of resources while respecting a provided
throughput and latency SLO~\cite{nexus}. It determines the right batch size for each model
while obeying latency and throughput constraints
and simultaneously minimizing the number of resources used. 
InferLine uses adaptive batching to determine how best to satisfy latency SLOs for pipelines consisting of one or
more ML models~\cite{crankshaw2020inferline}, and others have use adaptive batching to support inference with serveless systems~\cite{adaptive-batching}. 
NVIDIA Triton and \torchserve~\cite{triton, torchserve} supports adaptive batching and
concurrent model execution (i.e., multiple models can be served on the same GPU for better memory and compute utilization).

\paragraph{Configuration optimizations.}
McBench~\cite{mcbench} uses internal model knowledge and inputs to
generate TensorFlow configurations.
TensorTuner~\cite{tensortuner} applies gradient-based optimization to TensorFlow's threading; however, it does not partition an input batch across instances, support reconfiguration, or target optimizing inference latency in the context of serving systems.
TVM~\cite{tvm} and Tensor Comprehension~\cite{tensor-comprehensions} are compilers that generate optimized code for operators on a variety of hardware backends.
ParaX~\cite{parax} advocates for single-threaded inference to avoid memory and CPU stalls.

\paragraph{CPU-based inference performance.} Many optimizations have helped reduce inference latency of popular models. For example, removing zero padding and using tensors with dynamic shapes, using optimized matrix multiplication libraries, fusing memory-bound operators to increase arithmetic intensity, and effectively using vector units~\cite{bert-optimizing,bert-optimizing1,bert-optimizing2, slide}.
Many such techniques are part of optimized libraries such as Intel's MKL that are used in PyTorch and \torchserve, and they help bolster the baseline that \sys{} improves over.

\paragraph{Production serving systems.}
Amazon SageMaker, Azure ML, and TensorFlow Serving~\cite{sagemaker, azureml,tensorflow_serving} are serving systems with extra functions for serving ML models, including making it easy to manage and serve various model versions with
minimal overhead. However, these systems do not consider how models should be partitioned over
a multicore server, and the effect of \itbitb{} configuration on end-to-end latency.

\section{Discussion}
\label{sec:discussion}

\paragraph*{Storing model weights for large number of instances.} \sys converts a single
fat instance into multiple thin(-er) instances; as a result, model weights are duplicated in memory, once for each instance.
To combat this, we ported Ray's~\cite{ray} shared-memory store to \torchserve~\cite{model-loading}, which allowed instances to share the same model weights, saving memory.
Despite memory savings, this had negligible impact on performance, so all of the experiments in this paper do not use the shared-memory model store.
However, it can be useful for reducing the memory footprint, especially for
large models or large numbers of instances.
We leave this as future work; in any case,
the \torchserve community is working on its own shared-memory model store.

\paragraph{Newer CPU architectures.} \sys is designed to work with the current
generation of CPUs. 
Future CPUs will have more optimizations for deep learning
workloads. For example, the Sapphire Rapids CPU will have AVX512 VNNI, AMX, and
AVX512 BF16 optimizations to improve the compute performance of DNN workloads.
Sapphire Rapids will also have High Bandwidth Memory (HBM) to improve the memory
bandwidth.
These memories are based on HBM2E. The theoretical bandwidth for HBM2E-based memory
is expected to be 410 GB/s per 16~GB stack and Sapphire Rapids will have up to
4 stacks per CPU resulting in 1.6 TB/s of bandwidth and 64~GB capacity. This
makes the CPU memory bandwidth comparable to the GPU memory bandwidth,  with
capacity much higher than the typical GPU memory capacity.

Unsurprisingly, these improvements (to floating point performance and memory bandwidth) correspond to the bottlenecks we encountered.
It will be interesting to see how these improvements impact \sys{}.
Since CPU architectural and micro-architectural details are a black-box to \sys{}, it should work just as well for these CPUs.
License-based downclocking and memory bandwidth explained the differences between \sys{}'s expected and actual speedups, so it may be the case that its expected and actual speedups are closer on Sapphire Rapids CPUs.

\paragraph{NUMA awareness.} Parallelizing inference workload across non-uniform memory access~(NUMA) domains can negatively affect performance; hence, the prevailing sentiment is to avoid spanning NUMA domains~\cite{hyperthreading}. \sys{}'s resource allocator~(\S{\ref{sec:resource_allocator}}) places each instance entirely within a single socket whenever possible; however, in the worst case \textit{at most} one instance spans sockets.
The instance that spans across sockets would suffer from higher memory latencies due to remote NUMA accesses. One solution might be to replicate model weights on each node, and each thread only accesses node-local weights. We leave this work for the future.

\section{Conclusion}
\label{sec:conclusion}

Optimizing CPU-based inference to minimize latency for a given workload is challenging.
Using all of the threads for intra-op parallelism rarely results in optimal latency and neither does using all of the threads for multi-instance parallelism.
Also, picking the right configuration depends on the model and the CPU hardware. 
\sys{} solves this using an automated approach that combines selective profiling, an optimizer that estimates the performance of unprofiled configurations and suggests configurations to minimize latency, and performs online reconfigurations to avoid serving downtime. 
Collectively, these let \sys{} realize latency and throughput speedups of 1.43$\times$~to~1.83$\times$ averaged across batch sizes
on a range of common DNNs.

\section*{Acknowledgments}
Ankit Bhardwaj contributed to this work as a PhD student at University of Utah
and during internship at Microsoft Research.  This material is based upon work
supported by the National Science Foundation under Grant No.\ CNS-1750558. Any
opinions, findings, and conclusions or recommendations expressed in this
material are those of the authors and do not necessarily reflect the views of
the National Science Foundation.

\balance
\bibliographystyle{ACM-Reference-Format}
\bibliography{main}

\end{document}